\documentclass[preprint2]{aastex}

\slugcomment{Submitted to ApJ on 2006 Jan 19}

\lefthead{Tomita et al.}
\righthead{The Light Curves of SN2002ap}

\begin{document}


\title{The Optical/Near-Infrared Light Curves of SN~2002ap
for the First 1.5 Years after Discovery}

\author{Hiroyuki Tomita\altaffilmark{2},
    Jinsong Deng\altaffilmark{1,3},
        Keiichi Maeda\altaffilmark{4},
   Yuzuru Yoshii\altaffilmark{2,5},
      Ken'ichi Nomoto\altaffilmark{3,5},
        Paolo A. Mazzali\altaffilmark{3,6,7},
      Tomoharu Suzuki\altaffilmark{3}
   Yukiyasu Kobayashi\altaffilmark{8},
       Takeo Minezaki\altaffilmark{2},
         Tsutomu Aoki\altaffilmark{2},
           Keigo Enya\altaffilmark{9},
and   Masahiro Suganuma\altaffilmark{8} }

\altaffiltext{1}{Correspondence author's address:
                 National Astronomical Observatories,
                 Chinese Academy of Sciences,
                 20A Datun Road, Chaoyang District,
                 Beijing 100012, China; jsdeng@bao.ac.cn}
\altaffiltext{2}{Institute of Astronomy, School of Science,
                 University of Tokyo, 2-21-1
                 Osawa, Mitaka, Tokyo 181-0015, Japan}
\altaffiltext{3}{Department of Astronomy, School of Science,
                 University of Tokyo, 7-3-1
                 Hongo, Bunkyo-ku, Tokyo 113-0033, Japan}
\altaffiltext{4}{Department of Earth Science and Astronomy,
                 Graduate School of Arts and Science,
                 University of Tokyo, Meguro-ku,
                 Tokyo 153-8902, Japan}
\altaffiltext{5}{Research Center for the Early Universe,
                 School of Science, University of Tokyo, 7-3-1
                 Hongo, Bunkyo-ku, Tokyo 113-0033, Japan}
\altaffiltext{6}{Osservatorio Astronomico, Via Tiepolo 11,
                 34131 Trieste, Italy}
\altaffiltext{7}{Max-Planck-Institut f\"{u}r Astrophysik,
                 Karl-Schwarzschild-Strasse 1,
                 85748 Garching, Germany}
\altaffiltext{8}{National Astronomical Observatory, 2-21-1
                 Osawa, Mitaka, Tokyo 181-8588, Japan}
\altaffiltext{9}{Institute of Space and Astronomical Science,
                 Japan Aerospace Exploration Agency, 3-1-1,
                 Yoshinodai, Sagamihara, Kanagawa 229-8510, Japan}


\begin{abstract}

Late-time $BVRIJHK$ photometry of the peculiar Type Ic SN 2002ap,
taken between 2002 June 12 and 2003 August 29 with the MAGNUM
telescope, is presented. The light curve decline rate is derived
in each band and the color evolution is studied through comparison
with nebular spectra and with SN 1998bw. Using the photometry, the
$OIR$ bolometric light curve is built, extending from before light
maximum to day 580 after explosion. The light curve has a
late-time shape strikingly similar to that of the hypernova SN
1998bw. The decline rate changes from 0.018 mag/day between day
130 and 230 to 0.014 mag/day between day 270 and 580. To reproduce
the late-time light curve, a dense core must be added to the 1-D
hypernova model that best fits the early-time observations,
bringing the ejecta mass from 2.5 $M_\sun$ to 3 $M_\sun$ without
much change in the kinetic energy, which is $4\times 10^{51}$
ergs. This is similar to the case of other hypernovae and suggests
asymmetry. A large $H$-band bump developed in the spectral energy
distribution after $\sim$ day 300, probably caused by strong [Si
I] 1.646$\mu$m and 1.608$\mu$m emissions. The near-infrared flux
contribution increased simultaneously from $<30\%$ to $>50\%$ at
day 580. The near-infrared light curves were compared with those
of other Type Ib/c supernovae, among which SN 1983I seems similar
to SN 2002ap both in the near-infrared and in the optical.

\end{abstract}
\keywords{supernovae: general---supernovae: individual (SN
2002ap)---supernovae: photometry}


\section{Introduction}

There is growing interest in Type Ib/c supernovae (SNe) following
the discovery of a subclass that display very broad-lined spectra,
indicating the existence of ejecta expanding with velocity
$>0.1c$. Models of the optical spectra and the light curves (LCs)
of this high-velocity subclass, also called hypernovae (HNe, e.g.,
\citealt{nom04}), have concluded that these are the explosions of
massive C+O stars, producing up to 10 times the kinetic energy of
normal core-collapse SNe (e.g.,
\citealt{iwa98,iwa00,maz03,den05}). It is unclear why some HNe are
apparently associated with GRBs, i.e., SN 1998bw and GRB 980425,
\citep{gal98}, SN 2003dh and GRB 030329 \citep{sta03}, and SN
2003lw and GRB 031203 \citep{mal03}, while others are not.

The Type Ic SN 2002ap, discovered in M74 \citep{nak02}, is one of
the nearest SNe in recent years. Its host galaxy has a distance
modulus of only $29.5_{-0.2}^{+0.1}$ (\citealt{sha96,soh96}; see
also \citealt{hen05} for the latest review). \citet{tak02}
measured the total extinction toward the SN with a high-resolution
spectrum and concluded that $E(B-V)=0.09\pm 0.01$. We adopt these
values in this paper.

Early-time optical spectroscopy and photometry was published by
\citet{maz02}, \citet{gal02}, \citet{kin02}, \citet{fol03},
\citet{pad03a}, \citet{vin04}, and others. Most authors emphasized
the spectral similarity to the broad-lined SN 1998bw and SN
1997ef. The early-time light curve (LC), was much broader than
that of the normal Type Ic SN 1994I, but was significantly fainter
than that of SN 1998bw and peaked earlier.

\citet{yos03} (hereafter Paper I) presented early-time
near-infrared (NIR) and optical photometry, obtained with the
MAGNUM telescope. Using the $UBVRIJHK$ photometry, they built a
well-sampled $OIR$ bolometric LC for SN 2002ap. NIR photometry has
also been reported by \citet{nis02} and \citet{mat02}.
\citet{ger04} described the evolution of NIR spectra, which were
dominated by lines of intermediate-mass elements (see also
\citealt{dan02}).

\citet{maz02} modelled the early-time optical spectra and
bolometric LC using 1-D codes. They concluded that SN 2002ap was
the energetic explosion of a C+O star, having evolved from a star
of 20-25 $M_\sun$ on the main sequence, and that it ejected $\sim$
2.5 $M_\sun$ material with a kinetic energy of $\sim$
$4\times10^{51}$ ergs, including $\sim$ 0.07 $M_\sun$ $^{56}$Ni.
This places SN 2002ap at the low-mass low-energy end of
hypernovae. They also constrained the explosion date to MJD
52300.0 $\pm$ 0.5 days. We use the explosion date as the reference
point for any SN epoch in this paper. Based on observed limits in
pre-discovery images, \citet{sma02} argued that the progenitor was
either a single W-R star of main-sequence mass $<40 M_\sun$ or
more likely part of an interacting binary.

SN 2002ap was also monitored in the radio \citep{ber02,sod05},
X-rays \citep{sut03,sor04}, and UV \citep{sor02}. \citet{ber02}
estimated quite a small energy in any relativistic material from
the faintness of the radio. They addressed the material with
velocities larger than the shock velocity ($\sim0.3c$). Therefore,
their results do not contradict the high energy obtained in the
1-D model for the bulk SN ejecta by \citet{maz02}, whose ejecta
velocity distribution has a cut-off at 70,000 km~s$^{-1}$. As a
possible confirmation of this result, the shock velocity derived
by \citet{bjo04} from models of the radio and X-ray observations
was also $\sim$ 70,000 km~s$^{-1}$.

SN 2002ap was not apparently associated with a GRB \citep{hur02}.
However, \citet{kaw02} discovered a polarized spectrum component
that resembled the total flux spectrum but redshifted by
$z\sim0.3$ (see also \citealt{leo02,wan03}). They proposed that
this component is due to scattering from jet material.
\citet{tot03} further suggested that the jet may be too
baryon-contaminated to trigger a GRB and that it was expanding
almost freely and hence was radio quiet. \citet{tom04} reproduced
a bump in the LC of the XRF 030723 afterglow with a SN 2002ap-like
LC.

SN 2002ap was monitored into the late nebular phase. \citet{fol03}
published optical photometry covering from 2002 June to December
and spectroscopy continuing to 2003 February. The spectra show
unusually strong and sharp-peaked [\ion{O}{1}] and \ion{Mg}{1}]
emission lines (see also \citealt{leo02}). Late-time spectroscopy
was also obtained by other observers (Y. Qiu; K. Kawabata; S. B.
Pandey; private communications), and late-time photometry by
\citet{vin04} and \citet{dor03}.

It is also important to have intensive late-time NIR observations
of Type Ib/c SNe. As demonstrated in the case of SN 1998bw
\citep{pat01,sol02}, the NIR contribution to the total flux is
quite large, in particular at late times. Late-time NIR
observations may also reveal molecule and dust formation, if any
exists (e.g., \citealt{ger04}). However, few Type Ib/c SNe have
been observed at late phases in the NIR. For SN 2002ap, the only
late-time NIR observation published so far is a spectrum taken in
2002 August \citep{ger04}.

In this paper, we report on $BVRIJHK$ imaging photometry obtained
for SN 2002ap between 2002 June 12 and 2003 August 29, completing
the 1.5 years follow-up project with the MAGNUM telescope (see
also Paper I). We describe the observations and data reduction and
present our multicolor LCs in \S 2. An $OIR$ bolometric LC is then
constructed using the photometry. In \S 3, we discuss the
bolometric LC, analyze the evolution of the spectral energy
distribution (SED), compare the late-time NIR LCs with those of
other Type Ib/c SNe, and present 1-D models for the bolometric LC.

\section{Imaging Photometry}

\subsection{Observations and Data Reduction}

Optical (Johnson-Cousins $BVRI$) and NIR ($JHK$) imaging
photometry of SN 2002ap ($\alpha_{2000}=01^h 36^m23.^s85$,
$\delta_{2000}=+15^{\circ}45'13.''2$) was carried out using the
MAGNUM 2m telescope at the summit of Haleakala on the Island of
Maui, Hawaii \citep{yos02}. The multicolor imaging photometer
mounted at the bent-Cassegrain focus has an SITe $1024\times 1024$
pixel CCD with a scale of $0.''277\;$pixel$^{-1}$, for which the
light is effectively received in a field of view of $430\times
430$ pixel CCD or $119''\times 119''$, and also has an SBRC
$256\times 256$ pixel InSb with a scale of $0.''346\;$pixel$^{-1}$
yielding a field of view of $88.''5\times 88.''5$.  A dichroic
beam splitter enables simultaneous imaging through optical
($UBVRI$) and NIR ($ZJHK'KL'$) filters \citep{kob98a}.

The observations were scheduled by the MAGNUM scheduler
\citep{kob98b}, except for the solar conjunction from March to
June in 2003.  Telescope dithering was performed with 6 or
10-arcsec steps primary for $JHK$.  The typical exposure time for
one step was 190 sec for $B$, 95 sec for $V$, 65 sec for each of
$RI$, 60 sec for $J$ and 30 sec for each of $HK$. As SN 2002ap
faded, the number of dithering steps was increased from 6 to 12.
The typical seeing size of the stellar image was 1.0-1.6 arcsec in
$V$ and 0.85-1.4 arcsec in $K$. Dome-flat images for $BVRIJHK$ and
NIR dark images were obtained at the end of each night.

Image reduction was performed using our package of IRAF-based
\footnote{IRAF is distributed by the National Optical Astronomy
Observatories, which are operated by the Association of
Universities for Research in Astronomy, Inc., under contract to
the National Science Foundation.} automated reduction software
(MAGRED), which includes the standard corrections for bias and
flat field frames. For more details see Paper I.

Two reference stars, A and B, were observed in the same frame of
SN 2002ap for differential photometry. One reference star (A,
$\alpha_{2000}=01^h 36^m19.^s49$,
$\delta_{2000}=+15^{\circ}45'20.''8$) was used for $BVRI$ as in
Paper I. The other reference star (B, $\alpha_{2000}=01^h
36^m20.^s32$, $\delta_{2000}=+15^{\circ}44'54.''3$) was used for
$JHK$. In order to calibrate the magnitude of star B, NIR standard
stars \citep{hun98} were also observed, whenever possible, either
before or after the observation of SN 2002ap. The median
magnitudes and the number of calibrations for star B in the InSb
frame are $J(6)=15.69\pm 0.02$, $H(6)=15.11\pm 0.05$, and
$K(6)=14.86\pm 0.03$.

A variability check was made for stars A and B by increasing the
total exposure time in proportion as SN 2002ap faded. Figure 1
shows the difference between their $BVRI$ magnitudes after 2003
July, when the optical flux of star B became detectable. (It was
not possible to make the same plot in the NIR because star A was
usually outside or on the edge of the InSb detector.) The error
bars estimated from photon statistics are in good agreement with
the dispersion of the data points, so that it is not necessary to
regard stars A and B as variable.

Simple aperture photometry was used throughout, and the aperture
magnitude of SN 2002ap was obtained as the magnitude difference
between SN 2002ap and star A or B.  An aperture size of 6.9 arcsec
was adopted for all images, which minimizes the dispersion of the
differential photometry.  However, special care was taken for
reliable magnitude determination of SN 2002ap, given the presence
of a faint background object within a distance of about 1 arcsec.
This object became distinguishable from SN 2002ap, when SN 2002ap
became sufficiently faint after solar conjunction in 2003. In 2003
December, SN 2002ap faded compared to the background object, and
it was easy to resolve SN~2002ap and the object separately. Then,
using the IRAF DAOPHOT package, the PSF magnitude of the object
was measured several times when the seeing was relatively good.
This way, the PSF magnitude of the object was determined as
$B=21.7\pm0.5$, $V=21.6\pm0.2$, $R=21.25\pm0.12$,
$I=20.62\pm0.17$, $J=20.0\pm0.1$, $H=19.3\pm0.1$, and
$K=19.1\pm0.3$. These fluxes were subtracted from each aperture
photometry of SN 2002ap.

Moreover, reference star B in NIR photometry has a faint nearby
star which is about 4 mag fainter. Photometry was performed on the
images where this faint nearby star was subtracted using IRAF
DAOPHOT package.

\subsection{Observed Light Curves}

The $BVRIJHK$ data taken over 42 nights from June 12, 2002 to Aug
29, 2003 are tabulated in Table 1.  The uncertainty in each
magnitude was derived combining in quadrature the magnitude
dispersion from the frames obtained during the same night and the
magnitude error of reference star A for $BVRI$, or B for $JHK$. A
small error of 0.01 mag was added in quadrature for the $BVRI$
magnitudes, taking into account the non-linearity correction error
between SN 2002ap and reference star A. Another 0.01mag error
caused by the seeing effect, arising from the profile difference
between SN 2002ap and the reference star A, was also added. The
absolute calibration error and the subtracted flux uncertainty of
the background object were taken into account when evaluating the
total photometry error. No color-term correction was applied to
the magnitudes of SN 2002ap in the nebular phase, because strong
line emissions and large color difference with respect to the main
sequence star make this correction particularly difficult, as
demonstrated by \citet{men89}.

We show the MAGNUM $UBVRIJHK$ LCs in Figure 2 ({\em filled
circles}), combining the late-time data in Table 1 and the
early-time data of Paper I. For comparison, we also show the
$UBVRI$ photometry of \citet{pad03a,pad03b} ($<340$ days; {\em
crosses}), that of \citet{fol03} ($<320$ days; {\em open
circles}), and the early-time $JHK$ photometry of \citet{nis02}
($<50$ days; {\em open circles}). Obviously, our LCs have the
largest time span. They are unique also in that both the early and
late phase were well covered in NIR, the first time for any SN
Ib/c. To guide the eye, we made B-spline fits ({\em solid lines})
to the data, excluding the Pandey and Foley late-time ones.

It is clear from the figure that the decline of the late-time LC
slowed down after September 2002 in several bands. We derived the
decline rates and the date when the decline rate changed in each
band by $\chi^2$-fitting the late-time MAGNUM data using a broken
line (see Table 2). Before the change, which took place between
about day 210 and 250, the decline rate was similar in all bands,
i.e., about $0.016-0.022$ mag/day. The change is most profound in
the NIR, as the $J$ and $H$ LCs flattened from 0.018 mag/day to
0.012 and 0.008 mag/day, respectively. It is small for most
optical bands, although the $V$ LC did flatten from 0.021 to 0.014
mag/day. As a result, the NIR contribution to the total flux
increased significantly going to very late times. This will be
discussed further in the next section.

Most of our late-time data agree well with the \citet{fol03} ones,
although their $B$ magnitudes are 0.1-0.2 mag brighter than ours
after day 280 and their $I$ magnitudes 0.1-0.25 mag brighter after
day 160. The \citet{pad03b} late-time data are 0.1-0.15 mag
brighter than ours in the $B$ band, 0.2-0.4 mag brighter in the
$V$ band, and 0.1-0.2 mag fainter in the $I$ band. Those
differences can be partly explained by measurement uncertainties
inherent to practical broadband photometry of SNe (e.g., see
\citealt{ham90}), since these objects have time-evolving and
line-dominated spectra of non-stellar nature. Our late-time $BVRI$
LCs can differ from the those of \citet{vin04} and \citet{dor03}
by as much as 0.3 mag. However, their data were less well sampled
than ours.

The late-time evolution of the optical colors, $B-V$, $V-R$, and
$R-I$, is very different between SN 2002ap and SN 1998bw, as shown
in Figure 3. In particular, $V-R$ of SN 2002ap between day 140 and
330 is about 0.3-0.6 mag larger and $B-V$ between day 200 and 340
about 0.3 mag smaller than in SN 1998bw. This can be easily
understood within the context of spectroscopy. In the nebular
phase, the spectra of both SNe were dominated by [\ion{O}{1}]
$\lambda\lambda$6300,6364 in the $R$ band, [\ion{Fe}{2}]
multiplets in the $V$ and $B$ bands, and \ion{Mg}{1}]
$\lambda$4571 in the $B$ band, while \ion{Ca}{2}]
$\lambda\lambda$7291,7324 may also contribute to the $R$ and $I$
photometry and the \ion{Ca}{2} NIR triplet to the $I$ photometry
depending on the actual filter cut-off. As discussed by
\citet{fol03}, SN 2002ap has the strongest [\ion{O}{1}]
$\lambda\lambda$6300,6364 and \ion{Mg}{1}] $\lambda$4571 nebular
emissions of any SN published. However, its [\ion{Fe}{2}]
multiplets are much weaker than in SN 1998bw, a consequence of
much less $^{56}$Ni ejected by SN 2002ap. Between day 390 and 520,
$V-R$ and $R-I$ dropped and $B-V$ rose rapidly. This may indicate
a flux shift from emissions of intermediate-mass elements to
[\ion{Fe}{2}] multiplets. Unfortunately, no observed spectra have
so far been obtained for Type Ib/c SNe at such late times.

\section{Discussion}

\subsection{Bolometric Light Curve}

We built the $OIR$ bolometric LC integrating the $BVRIJHK$
broadband flux (see Paper I for details). A reddening of
$E(B-V)=0.09\pm0.01$ was corrected for \citep{tak02}, and a
distance modulus of 29.5 was adopted (but see \citealt{hen05}).

The LC is shown in Figure 4, extending that in Paper I to $\sim
580$ days since explosion. It consists of three independent data
sets, which agree with one another. Most of the data ({\em filled
circles}), tabulated in Table 3 ({\em late time}) and Paper I
({\em early time}), were made from the MAGNUM photometry. We
interpolated and extrapolated the observations to estimate the the
$K$ magnitudes between day 280 and 340 and the $J$ and $I$
magnitudes at day 520. For epochs $<11$ days, the observations of
\citet{mat02}, \citet{gal02}, and \citet{coo02} were used for
interpolation. Bolometric magnitudes at day 407 and 578 ({\em
triangles}) were converted from $H$ magnitudes using approximate
bolometric corrections. The other two data sets were based on the
\citet{fol03} ({\em open circles}) and \citet{pad03a,pad03b} ({\em
crosses}) optical photometry, respectively, which was combined
with the \citet{nis02} and our $JHK$ photometry.

The bolometric LC at late time has a shape similar to that of SN
1998bw (Figure 4; {\em squares}), another hypernova, although the
latter was much brighter. The SN 1998bw LC in Figure 4 is shifted
down by 1.75 mag to match the peak magnitude of SN 2002ap. Both
LCs show a similar late-time slowing-down. For SN 2002ap, the
decline rate changed from $\sim 0.018$ mag/day between day 130 and
230 to $\sim 0.014$ mag/day between day 270 and 580. For SN
1998bw, the decline rate was $\sim 0.019$ mag/day between day 70
and 220 and $\sim 0.014$ mag/day between day 320 and 540. For
comparison, the late-time bolometric LC of another hypernova, SN
1997ef, followed the decay rate of $^{56}$Co, i.e., $\sim 0.01$
mag/day \citep{maz04}. On the other hand, SN 2002ap reached its
light maximum $\sim 5$ days earlier than SN 1998bw did, although
their peak widths are actually comparable (Figure 4; {\em inset}).

\subsection{Spectral Energy Distribution}

We show the SED of SN 2002ap at 5 typical late epochs in Figure 5,
where the monochromatic fluxes converted from the MAGNUM $BVRIJHK$
photometry are connected using spline-fitting curves. The observed
upper limits of the $K$-band flux at day 393 and 520 are marked by
{\em arrows}. Zero flux was assumed at both the the blue edge of
the $B$ band and the red edge of the $K$ band. We used such SEDs
to obtain the $OIR$ bolometric magnitudes.

The SEDs are dominated by the flux in the $R$ and $I$ bands before
about day 340-390, which was mainly due to strong [\ion{O}{1}],
\ion{Ca}{2}], and \ion{Ca}{2} emissions \citep{fol03}. The $B$
bump at the three intermediate epochs reflects the contribution of
\ion{Mg}{1}] $\lambda$4571. After about day 340-390, the $V$ flux,
attributed to [\ion{Fe}{2}] multiplets, increased relative to that
in other optical bands, while the $I$-band-dominated \ion{Ca}{2}
emissions died away.

The NIR flux contribution rose rapidly after $\sim$ day 300
(Figure 6). This can been seen in Figure 5 as a big $H$ bump in
the SED and also in Figure 2 as a significant flattening of the
$H$ and $J$ LCs. To explain this, we examined the synthesized NIR
spectra of the models that reproduce the late-time optical spectra
of SN 2002ap (P. A. Mazzali et al. 2005, in preparation). The most
likely candidates are the strong [\ion{Si}{1}] 1.646 $\mu$m and
1.608 $\mu$m lines, while the 1.099 $\mu$m line may account for
the concurrent $J$-band flux increase. The lines were already
developed in the day $\sim$ 200 NIR spectrum of \citet{ger04},
although they were still not as strong as a 1.5 $\mu$m feature,
possibly \ion{Mg}{1} emission. We note in passing that the
appearance of the CO first-overtone band in that spectrum
suggested by those authors seems to coincide with the $K$-band LC
steepening seen in Figure 2.

The late-time $H$-band spectra of SN 1983N, taken by \citet{gra86}
and of very low resolution, showed a strong 1.65 $\mu$m feature,
which was explained by those authors as [\ion{Fe}{2}] lines but
was soon, together with an adjacent feature, re-identified as
[\ion{Si}{1}] lines by \citet{oli87}. These [\ion{Si}{1}] NIR
features can also be found in the synthetic late-time spectra of
Type Ib SNe computed by \citet{fra89}.

\subsection{Late-time NIR Light Curves of Type Ib/c Supernovae}

In Figure 7 we compare the late-time $JHK$ LCs of SN 2002ap with
those of other Type Ib/c SNe (see Table 4). SN 1998bw and SN 1984L
were 1.5--2.5 mag brighter than all other SNe, indicating a large
mass of $^{56}$Ni ejected. The NIR LCs of SN 1984L were relatively
slow, as was its late-time optical LC which was modelled using
very massive ejecta but with normal kinetic energy
\citep{swa91,bar93}. The few points of SN 1998bw,when shifted down
by 1.8 mag ($J$), 1.5 mag ($H$), and 1.7 mag ($K$), respectively,
fall on the LCs of SN 2002ap. The LCs of SN 1983N have later and
broader peaks than those of SN 2002ap, but they become similar
between in the $J$-band and $H$-band day $\sim$ 200 and 350. On
the other hand, the $JHK$ data of SN 1983I and 1982R seem to
bridge the LC gap in SN 2002ap data between day 50 and day 140
(solar conjunction).

Figure 7 contradicts the picture that SNe 1983I, 1983N, and 1984L
have similar $JHK$ LCs \citep{eli85}. Type Ib/c SNe were first
established as a new class by those authors, who showed that these
SNe have similar $JHK$ LCs which are different from those of Type
Ia SNe [and also by \citet{whe85} through an analysis of the
optical spectra]. They artificially shifted the data of SN 1983I
and SN 1984L in both phase and brightness in order to match the
NIR LCs of SN 1983N. Their SN 1983I data were assigned epochs 20
days later than ours, and SN 1984L 14 days earlier. Our epoch
estimates are more reasonable because we based them on optical LCs
and spectra which are better observed and understood than the NIR
data.

\subsection{Was SN 1983I similar to SN 2002ap?}

Inspecting by eye the only published spectrum of SN 1983I
\citep{whe87}, we noticed that it resembles those of SN 2002ap
between 2002 February 16 and 22 \citep{kin02}. The similarity can
be seen not only in the overall shape and strongest line features,
but also in relatively weak ones, like those near $4900$\AA,
$5900$\AA, and $6200$\AA. The spectrum was taken on 1983 May 17, 5
days after discovery. We estimate the epoch of that spectrum as
$22\pm3$ days since explosion, through this comparison, and hence
derive an explosion date of 1983 April 25.

The LCs of SN 1983I seem to match those of SN 2002ap not only in
the NIR (Figure 7) but also in the optical (Figure 8), if our
estimates of the explosion date and the distance modulus are
correct. We adopt 31.0 as the distance modulus to the host galaxy
of SN 1983I, NGC 4051 in the Ursa Major cluster \citep{tsv85}.
This value is also close to the average of the distance modulus of
the Ursa Major cluster (31.4, \citealt{tul00}), and that
corresponding to its radial velocity corrected for the Local Group
infall onto Virgo (30.5, LEDA
\footnote{http://leda.univ-lyon1.fr/}; \citealt{pat97})

Therefore, SN 1983I could be a precedent of the unusual SN 2002ap,
which lies at the low-mass low-energy end of hypernovae
\citep{maz02}. However, its data are insufficient for us to make a
solid conclusion.

\subsection{Modelling the Late-time Light Curve}

The late-time bolometric LC of SN 2002ap declines more slowly than
that of the 1-dimensional (1-D) model of \citet{maz02} that best
reproduced the early-time spectra and LC (see Figure 9; {\em
dotted line}). This is expected, since the LC follows that of SN
1998bw closely in terms of the decline rate, as shown in Figure 4.
The best-fitting 1-D model for the early-time spectra and LC of SN
1998bw also fails to explain its late-time spectrum and LC
\citep{nak01,maz01}. To mimic the outcome of 2-dimensional
jet-induced explosion calculations \citep{mae02}, which applies to
SN 1998bw and may also apply to SN 2002ap, \citet{mae03}
introduced a dense core. That core absorbs $\gamma$-rays
efficiently at late time. Using this structure, the slow LC
decline of hypernovae was reproduced by those authors using a
Monte-Carlo radiative transfer code but with simplified physics.

We tested the dense-core scenario on SN 2002ap using our
sophisticated 1-D SN radiation hydrodynamical and $\gamma$-ray
transfer code \citep{iwa00,maz02,den05}. The best-fitting model
({\em solid line}) has $\sim 0.6 M_\sun$ ejecta below 3,000
km~s$^{-1}$, compared with 0.1 $M_\sun$ in the model without a
dense core, and $\sim 0.01 M_\sun$ low-velocity $^{56}$Ni.
Compared with \citet{maz02}, who modelled only the early-time
observations, the total ejecta mass has increased from 2.5
$M_\sun$ to 3 $M_\sun$, but with little change in the total
$^{56}$Ni mass (0.08 $M_\sun$) and the kinetic energy ($4\times
10^{51}$ ergs). A dense ejecta core is also required in 1-D
late-time spectrum models (P. A. Mazzali et al. 2006, in
preparation) in order to explain the observed sharp line cores of
[\ion{O}{1}] and \ion{Mg}{1}] \citep{leo02,fol03}.

A dense core cannot be formed in 1-D explosion simulations for
hypernovae \citep{nak01}, but it is a natural product of 2-D
jet-induced explosions \citep{mae02}. This strongly indicates
asymmetry in the SN 2002ap explosion, an intrinsic feature also
shared by other hypernovae. Evidence of asymmetry in SN 2002ap can
also be found in spectropolarimetry \citep{kaw02,leo02,wan03},
which shows an intrinsic continuum polarization varying around
0.5\%, corresponding to an asphericity of $\sim$ 10\% for the bulk
of the ejecta.

\acknowledgements

We thank S. B. Pandey, E. Nishihara, and R. J. Foley for the
electronic files of their photometric data. This work has been
supported in part by the Grant-in-Aid for Scientific Research
(16540229, 17030005, \& 17033002 for KN) and the 21st Century COE
program (QUEST) from the JSPS and MEXT in Japan, and by the
National Science Foundation under Grant No. PHY99-07949.



\newpage

\figcaption[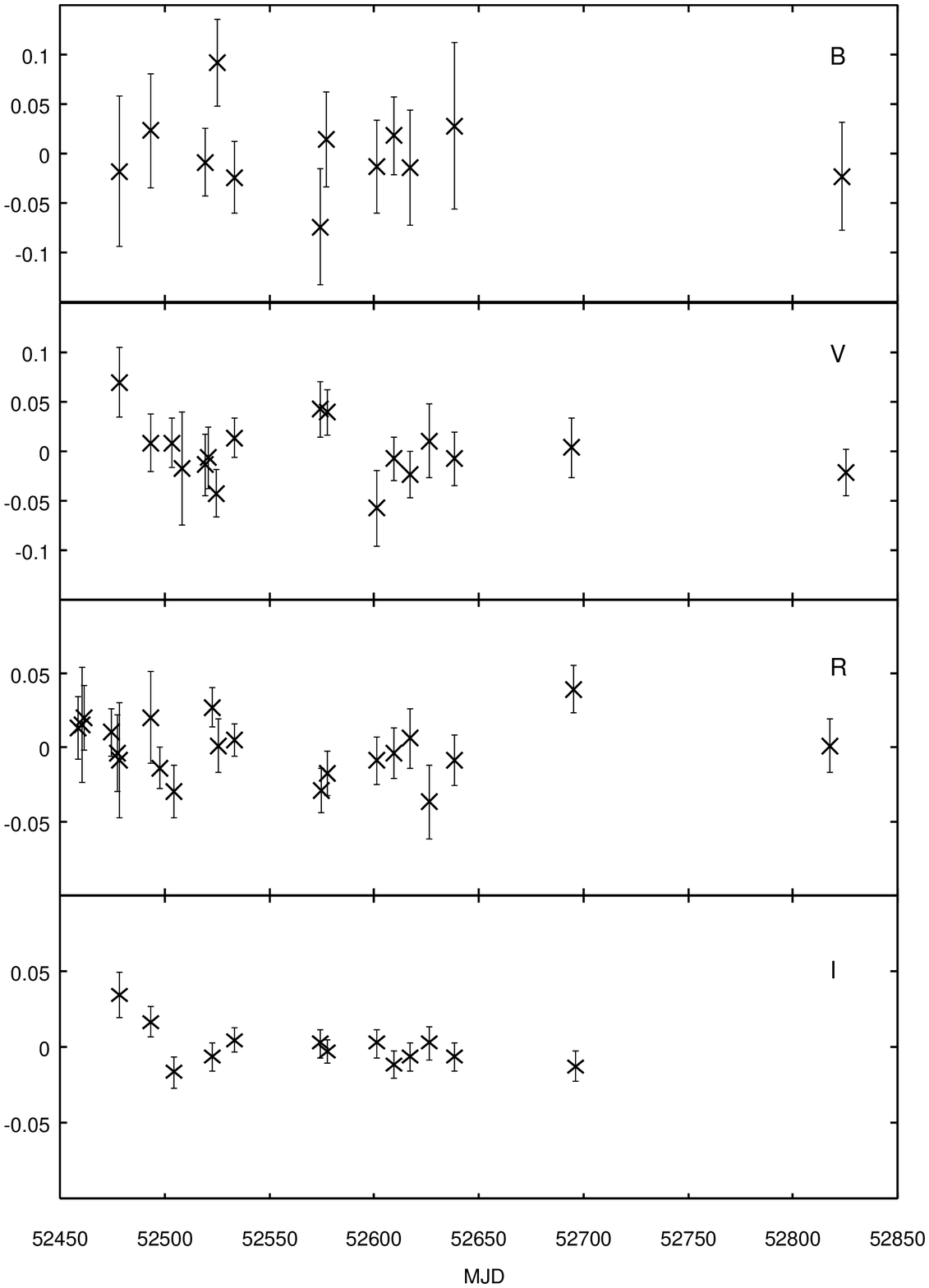]{ Difference between observed magnitudes of two
reference stars [$\Delta m\equiv m({\rm A})-m({\rm B})$] in the
$BVRI$ bands for MAGNUM observations after 2003 July. The mean
difference $\langle\Delta m\rangle$\ is shifted to zero. Error
bars are estimated from photon statistics. The standard deviation
of data distribution and mean error from photon statistics are
0.042/0.053 ($B$), 0.032/0.030 ($V$), 0.020/0.020 ($R$), and
0.014/0.010 ($I$). The computed reduced chi-square values were
0.67 for $B$ , 1.11 for $V$, 1.23 for $R$ and 1.33 for $I$.
\label{fig_refab} }

\figcaption[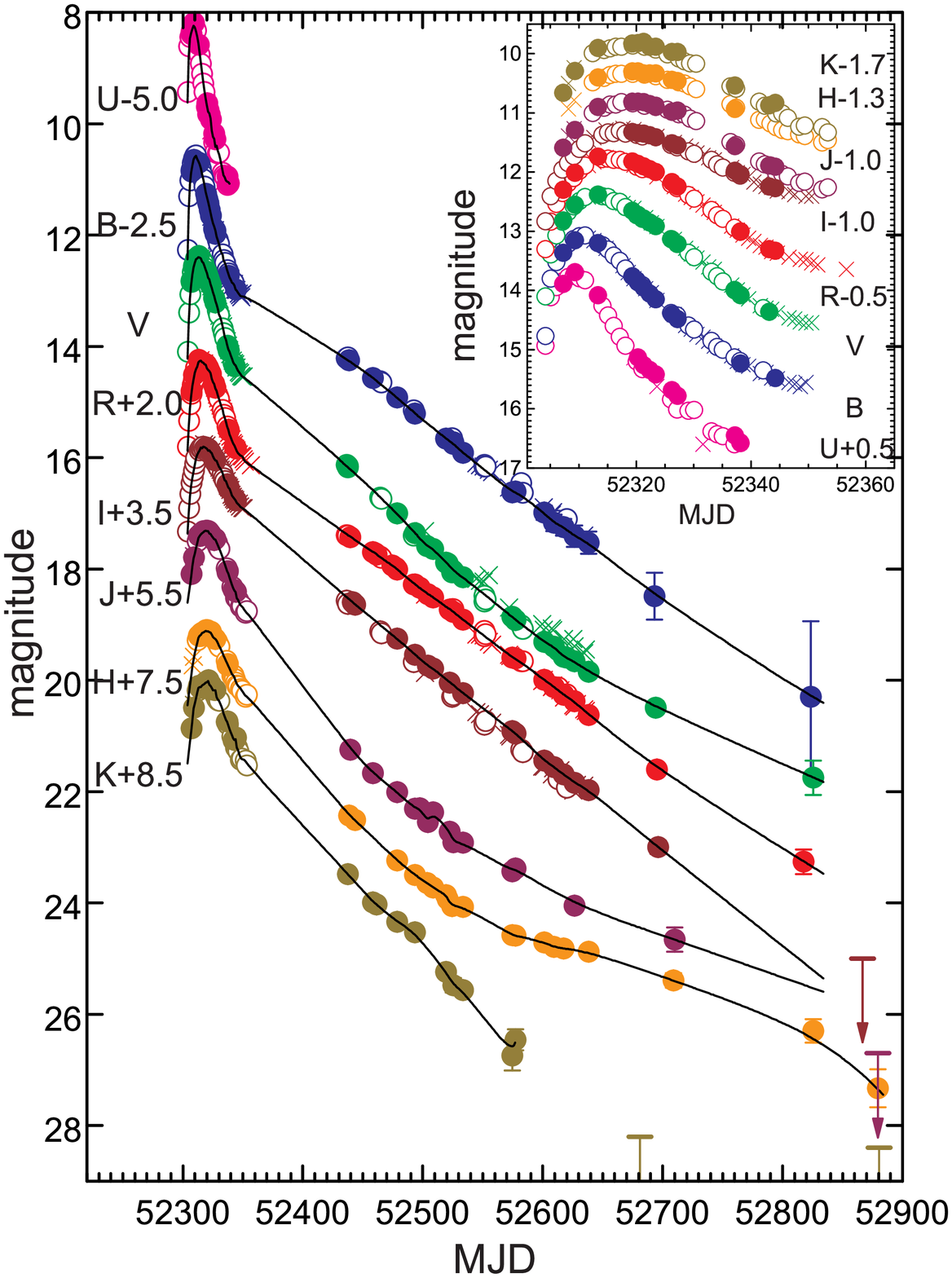]{$UBVRIJHK$ LCs of SN~2002ap for the first 1.5
years after discovery obtained with the MAGNUM telescope ({\em
filled circles}). Also shown are the $UBVRI$ data reported by
\citet{fol03} ({\em open circles}) and \citet{pad03a,pad03b} ({\em
crosses}) and the early-time $JHK$ data by \citet{nis02} (also
{\em open circles}). The LCs are shifted vertically in order to
avoid overlap. {\em Solid lines} are B-spline fittings to the
data, excluding the Pandey and Foley late-time ones. \label{fig
LC}}

\figcaption[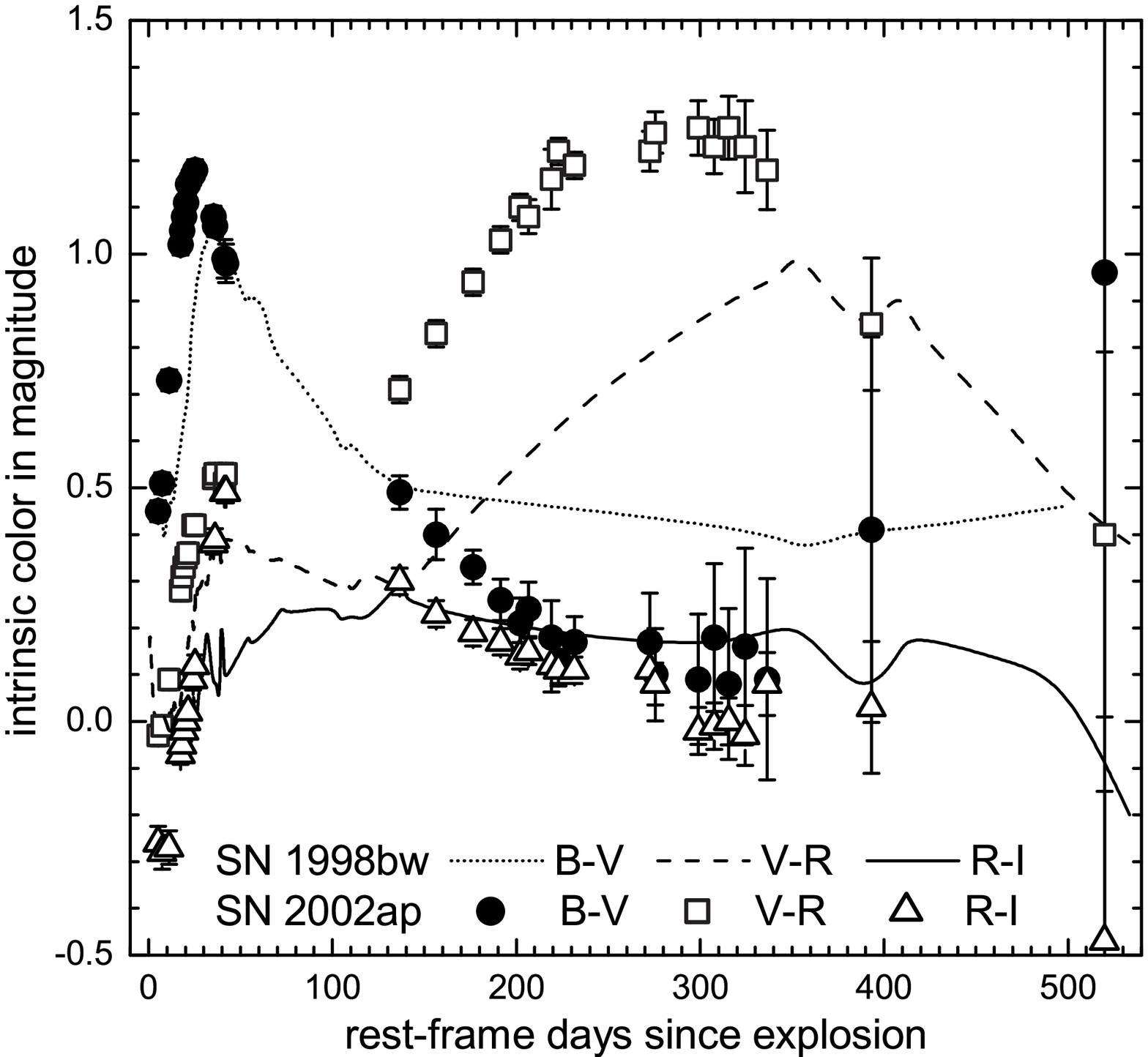]{Evolution of optical colors, $B-V$, $V-R$, and
$R-I$, of SN 2002ap compared with that of SN 1998bw. The colors of
SN 2002ap are calculated from the MAGNUM photometry, corrected for
$E(B-V)=0.09$, while those of SN 1998bw from the photometry in
\citet{gal98}, \citet{pat01}, and \citet{sol02}, corrected for
$A(V)=0.1$. \label{fig color}}

\figcaption[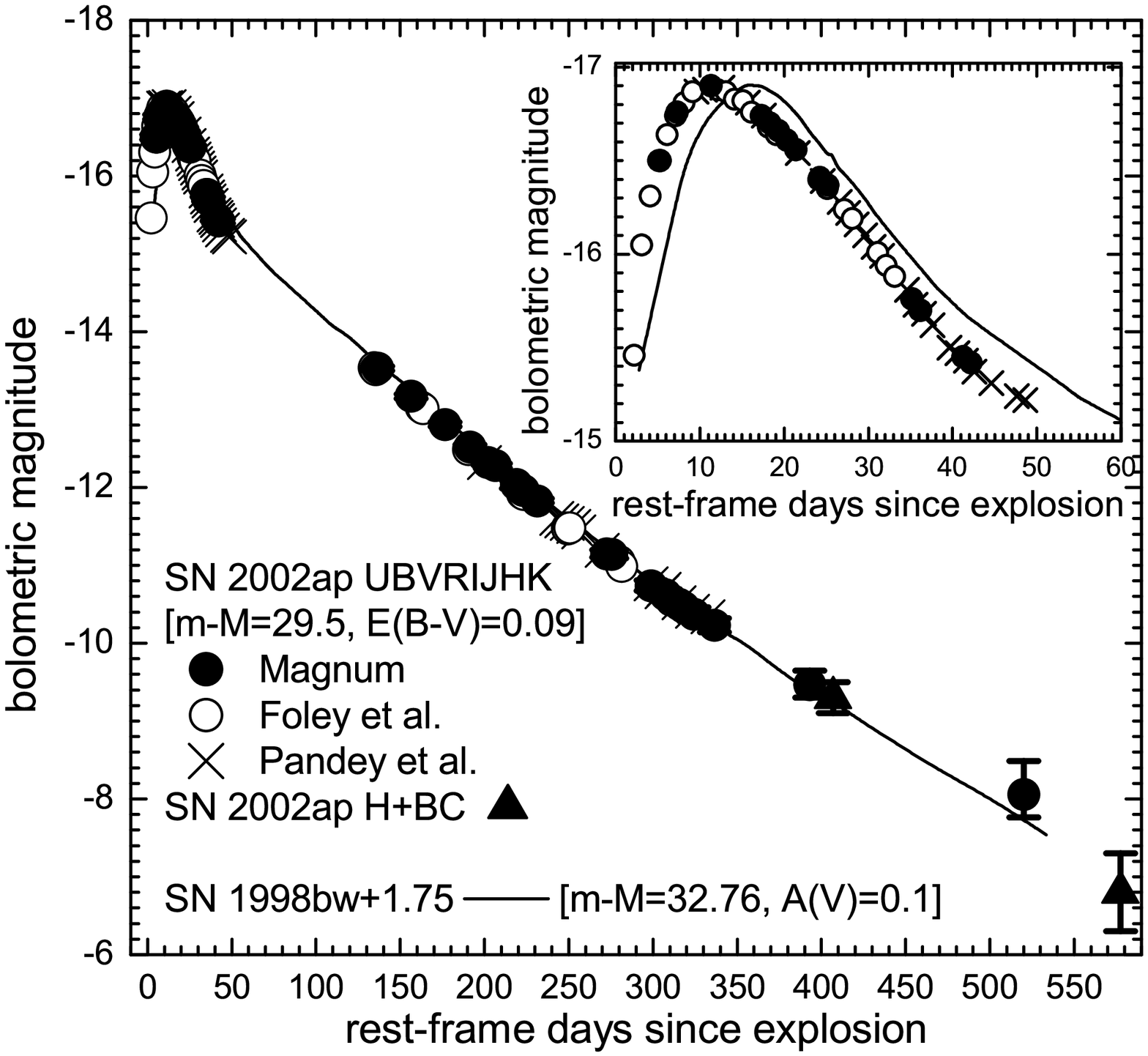]{$OIR$ bolometric LC of SN 2002ap constructed
using the MAGNUM $UBVRIJHK$ photometry, assuming $m-M=29.5$ and
$E(B-V)=0.09\pm 0.01$, compared with that of SN 1998bw
[$z=0.0085$, $m-M=32.76$, and $A(V)=0.1\pm 0.1$]. Also shown are
bolometric magnitudes based on the \citet{fol03} and
\citet{pad03a,pad03b} optical photometry, which are combined with
the MAGNUM and \citet{nis02} $JHK$ photometry, and tho ones
estimated from the MAGNUM $H$ photometry using approximate
bolometric corrections. The SN 1998bw bolometric LC is shifted
down by 1.75 mag to match the peak brightness of SN 2002ap. It is
constructed using the photometry in \citet{gal98}, \citet{pat01},
and \citet{sol02}. \label{fig bol}}

\figcaption[fig5.eps]{SEDs of SN 2002ap at 5 typical late epochs,
constructed using the MAGNUM photometry. The $BVRIJHK$
monochromatic flux is connected using spline-fitting curves. {\em
Capped arrows} represent flux upper limits in the $K$ band.
\label{fig sed}}

\figcaption[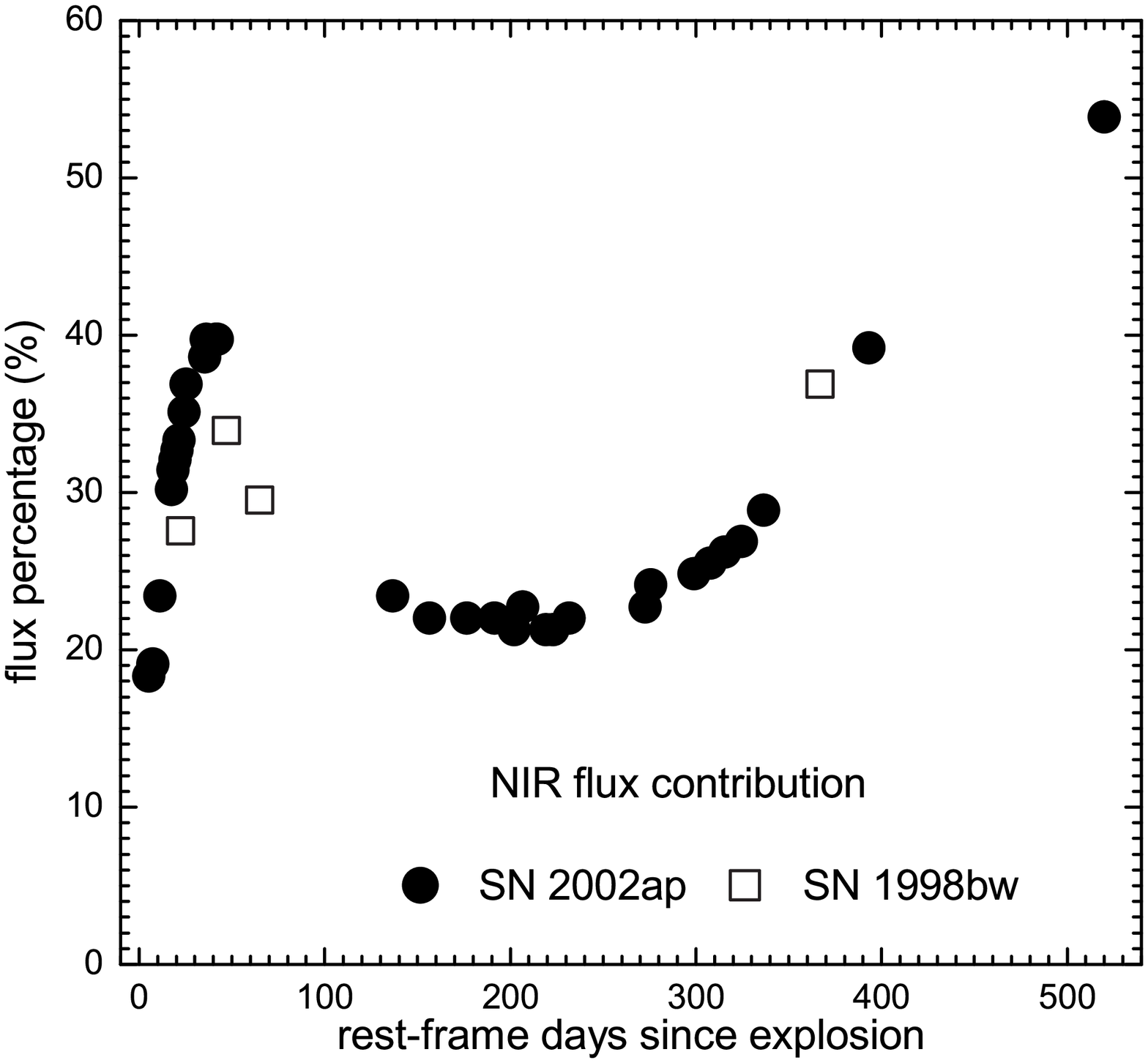]{Contribution of NIR flux to the total $OIR$
bolometric flux in SN 2002ap compared with that in SN 1998bw.
\label{fig fnir}}

\figcaption[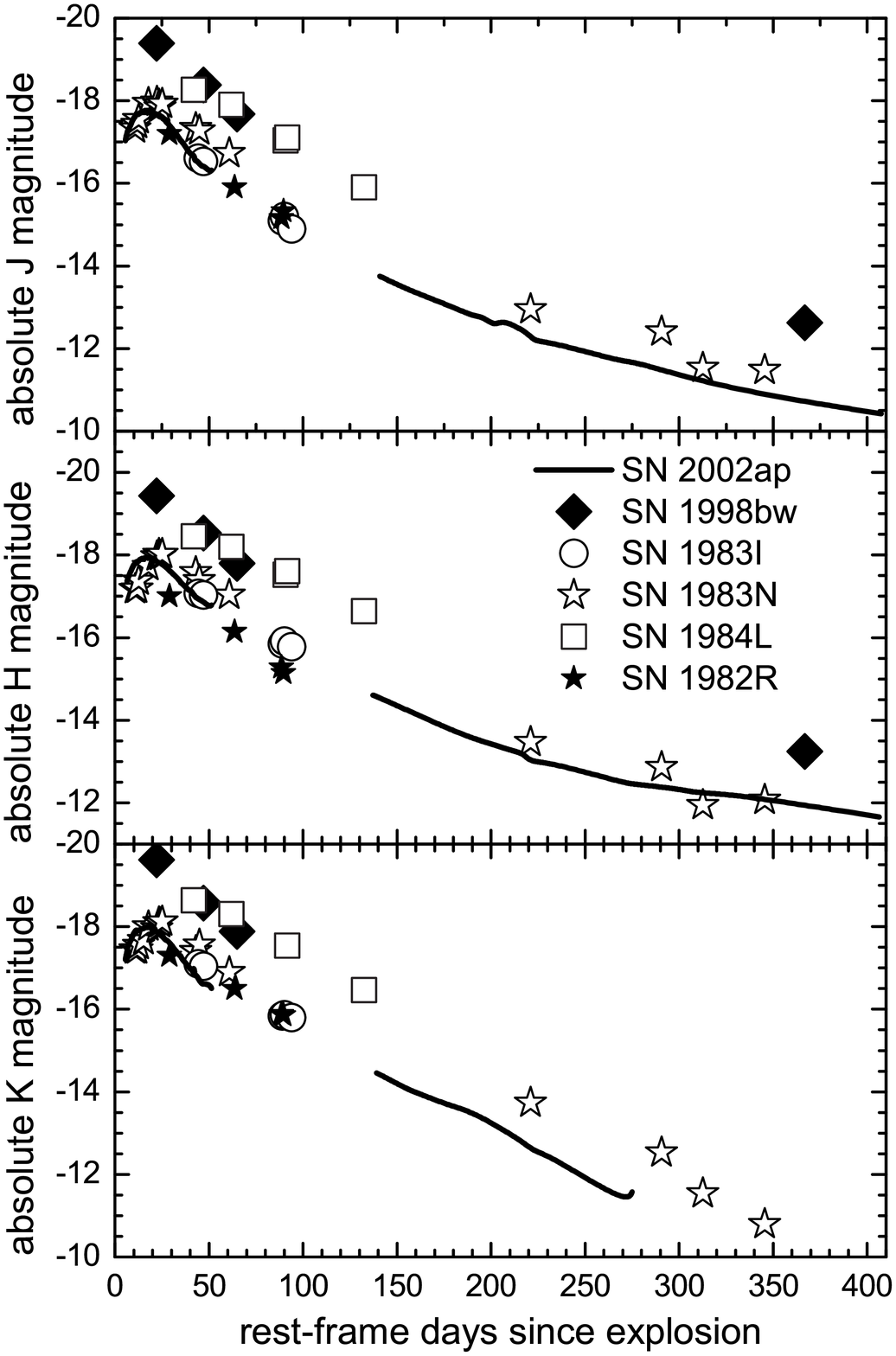]{Absolute $J$ ({\em top}), $H$ ({\em middle}),
and $K$ ({\em bottome}) LCs of the Type Ib/c SNe whose late-time
NIR photometry are available. See Table 4 for data references and
adopted SN parameters. \label{fig lcnir}}

\figcaption[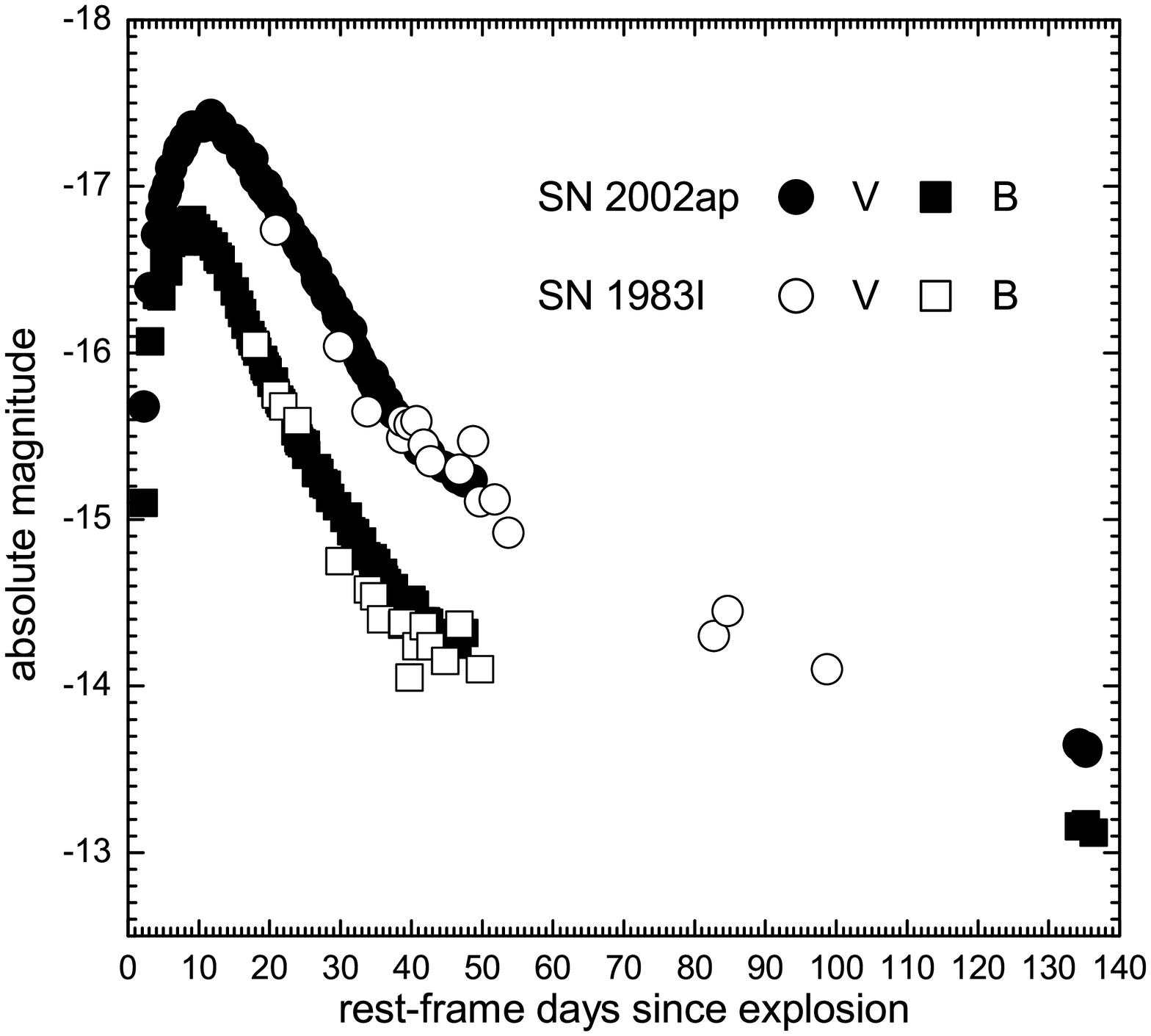]{Absolute $B$ and $V$ LCs of the Type Ic SN
1983I \citep{tsv85}, assuming $m-M\approx 31.0$ and an explosion
date of 1983 April 25 (see Table 4), compared with those of SN
2002ap.\label{fig 83I}}

\figcaption[f8.eps]{$OIR$ bolometric LC of SN 2002ap compared with
the best-fitting LCs of the 1-D ejecta model with a dense core at
$v<3,000$ km~s$^{-1}$ (total $M_{\rm ej}\approx 3$ $M_\sun$), and
that without a dense core (total $M_{\rm ej}\approx 2.5$
$M_\sun$). Both models have total kinetic energy of $4\times
10^{51}$ ergs and 0.08 $M_\sun$ $^{56}$Ni. \label{fig model}}


\onecolumn

\plotone{f1.eps}

\plotone{f2.eps}

\plotone{f3.eps}

\plotone{f4.eps}

\plotone{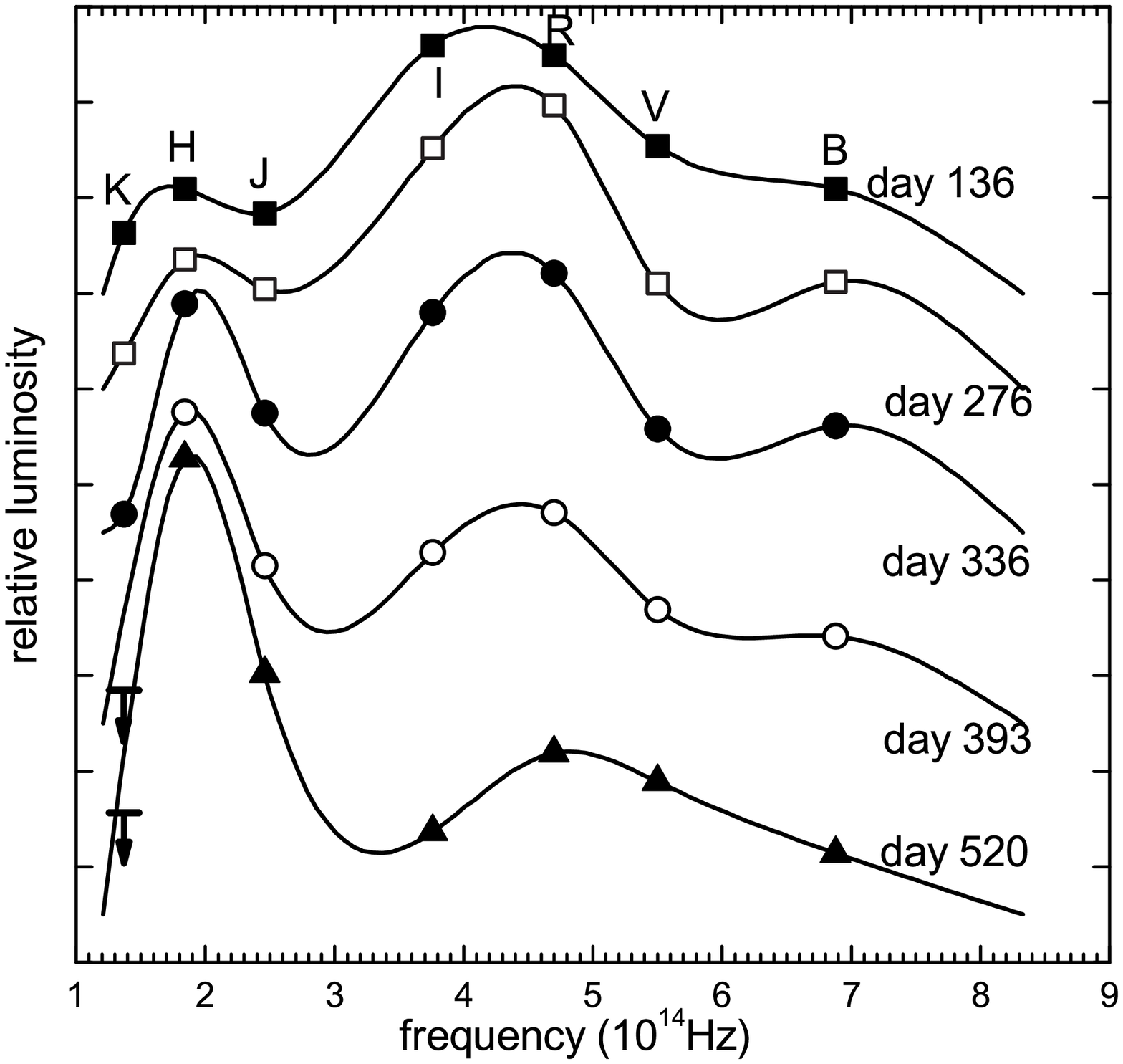}

\plotone{f6.eps}

\plotone{f7.eps}

\plotone{f8.eps}

\plotone{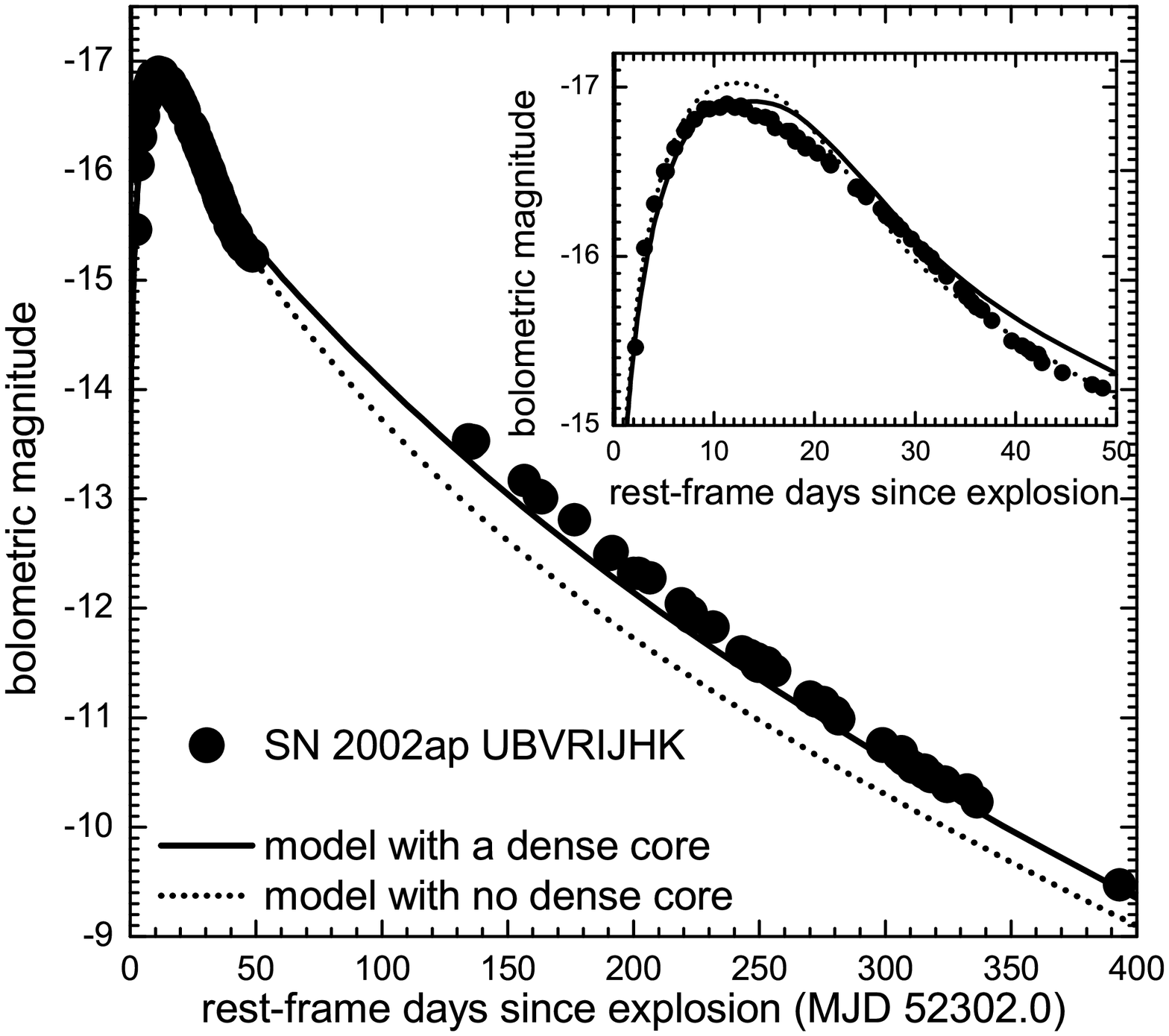}


{\scriptsize

\begin{deluxetable}{llcccccccc}
 \rotate
 \tabletypesize{\scriptsize}
 \tablecaption{Data of light curves of SN~2002ap}
 \tablewidth{0pt}
 \tablenum{1}
 \tablehead{
 \colhead{UT}     & \colhead{MJD}    & \colhead{$m_B$}  &
 \colhead{$m_V$}  & \colhead{$m_R$}  & \colhead{$m_I$}  &
 \colhead{$m_J$}  & \colhead{$m_H$}  & \colhead{$m_K$}  & Weather
}
 \startdata
 Jun 12.61  &  52437.61  & \nodata         & 16.15$\pm0.02$  & \nodata         & \nodata         & \nodata         & \nodata         & 14.98$\pm 0.04$ & clear \nl
 Jun 13.60  &  52438.60  & 16.75$\pm 0.03$ & \nodata         & \nodata         & \nodata         & \nodata         & 14.93$\pm 0.05$ & \nodata         & clear \nl
 Jun 14.60  &  52439.60  & \nodata         & \nodata         & 15.43$\pm 0.02$ & \nodata         & 15.75$\pm 0.03$ & \nodata         & \nodata         & clear \nl
 Jun 16.60  &  52441.60  & \nodata         & \nodata         & \nodata         & 15.10$\pm 0.02$ & \nodata         & \nodata         & \nodata         & clear \nl
 Jun 18.60  &  52443.60  & \nodata         & \nodata         & \nodata         & 15.14$\pm 0.02$ & \nodata         & 15.01$\pm 0.05$ & \nodata         & clear \nl
 Jul  3.56  &  52458.56  & 17.07$\pm 0.05$ & \nodata         & 15.70$\pm 0.02$ & \nodata         & 16.17$\pm 0.04$ & \nodata         & 15.49$\pm 0.06$ & thin cloud \nl
 Jul  5.60  &  52460.60  & \nodata         & \nodata         & 15.72$\pm 0.02$ & \nodata         & \nodata         & \nodata         & \nodata         & thin cloud \nl
 Jul  6.55  &  52461.55  & \nodata         & \nodata         & 15.74$\pm 0.02$ & \nodata         & \nodata         & \nodata         & 15.53$\pm 0.08$ & cloudy \nl
 Jul 19.56  &  52474.56  & \nodata         & \nodata         & 15.94$\pm 0.02$ & \nodata         & \nodata         & \nodata         & \nodata         & clear \nl
 Jul 22.56  &  52477.56  & \nodata         & \nodata         & 15.98$\pm 0.02$ & \nodata         & \nodata         & \nodata         & \nodata         & clear \nl
 Jul 23.51  &  52478.51  & 17.42$\pm 0.03$ & 17.00$\pm 0.02$ & 16.01$\pm 0.02$ & 15.75$\pm 0.02$ & 16.51$\pm 0.03$ & 15.74$\pm 0.06$ & 15.84$\pm 0.06$ & clear \nl
 Aug  7.47  &  52493.47  & 17.71$\pm 0.04$ & 17.36$\pm 0.02$ & 16.28$\pm 0.02$ & 16.04$\pm 0.02$ & 16.81$\pm 0.04$ & 16.00$\pm 0.07$ & 16.03$\pm 0.09$ & thin cloud \nl
 Aug 11.59  &  52497.59  & \nodata         & \nodata         & 16.34$\pm 0.02$ & \nodata         & 16.80$\pm 0.04$ & \nodata         & \nodata         & thin cloud \nl
 Aug 17.61  &  52503.61  & \nodata         & 17.59$\pm 0.02$ & \nodata         & \nodata         & \nodata         & 16.14$\pm 0.06$ & \nodata         & thin cloud \nl
 Aug 18.44  &  52504.44  & \nodata         & \nodata         & 16.45$\pm 0.02$ & 16.24$\pm 0.02$ & 17.05$\pm 0.04$ & \nodata         & \nodata         & clear \nl
 Aug 22.54  &  52508.54  & \nodata         & 17.64$\pm 0.03$ & 16.51$\pm 0.02$ & 16.29$\pm 0.02$ & 16.87$\pm 0.04$ & 16.22$\pm 0.07$ & \nodata         & cloudy \nl
 Sep  2.44  &  52519.44  & 18.16$\pm 0.05$ & 17.89$\pm 0.02$ & \nodata         & \nodata         & \nodata         & 16.36$\pm 0.07$ & 16.74$\pm 0.08$ & clear \nl
 Sep  3.61  &  52520.61  & \nodata         & 17.90$\pm 0.02$ & \nodata         & \nodata         & \nodata         & 16.43$\pm 0.07$ & \nodata         & thin cloud \nl
 Sep  5.59  &  52522.59  & \nodata         & \nodata         & 16.73$\pm 0.02$ & 16.54$\pm 0.02$ & 17.22$\pm 0.04$ & \nodata         & \nodata         & thin cloud \nl
 Sep  7.61  &  52524.61  & \nodata         & 18.02$\pm 0.02$ & \nodata         & \nodata         & \nodata         & 16.56$\pm 0.07$ & \nodata         & cloudy \nl
 Sep  8.41  &  52525.41  & 18.25$\pm 0.05$ & \nodata         & 16.77$\pm 0.02$ & \nodata         & 17.41$\pm 0.07$ & \nodata         & 16.98$\pm 0.13$ & cloudy \nl
 Sep 16.54  &  52533.54  & 18.41$\pm 0.05$ & 18.15$\pm 0.02$ & 16.91$\pm 0.02$ & 16.73$\pm 0.02$ & 17.42$\pm 0.04$ & 16.57$\pm 0.06$ & 17.07$\pm 0.10$ & clear \nl
 Oct 27.52  &  52574.52  & 19.12$\pm 0.10$ & 18.86$\pm 0.03$ & 17.59$\pm 0.03$ & 17.41$\pm 0.03$ & 17.93$\pm 0.07$ & 17.08$\pm 0.08$ & 18.24$\pm 0.27$ & clear \nl
 Oct 30.45  &  52577.45  & 19.11$\pm 0.09$ & 18.92$\pm 0.04$ & 17.61$\pm 0.02$ & 17.46$\pm 0.04$ & 17.88$\pm 0.06$ & 17.09$\pm 0.08$ & 17.96$\pm 0.19$ & thin cloud \nl
 Nov 23.22  &  52601.22  & 19.49$\pm 0.13$ & 19.31$\pm 0.05$ & 17.99$\pm 0.03$ & 17.94$\pm 0.04$ & \nodata         & 17.21$\pm 0.07$ & \nodata         & clear \nl
 Dec  1.39  &  52609.39  & 19.67$\pm 0.15$ & 19.40$\pm 0.05$ & 18.12$\pm 0.03$ & 18.06$\pm 0.04$ & \nodata         & 17.29$\pm 0.07$ & \nodata         & clear \nl
 Dec  9.35  &  52617.35  & 19.75$\pm 0.15$ & 19.58$\pm 0.06$ & 18.26$\pm 0.03$ & 18.19$\pm 0.04$ & \nodata         & 17.32$\pm 0.07$ & \nodata         & clear \nl
 Dec 18.37  &  52626.37  & 19.91$\pm 0.19$ & 19.66$\pm 0.09$ & 18.38$\pm 0.04$ & 18.34$\pm 0.05$ & 18.55$\pm 0.08$ & \nodata         & \nodata         & clear \nl
 Dec 30.37  &  52638.37  & 20.03$\pm 0.20$ & 19.85$\pm 0.08$ & 18.62$\pm 0.03$ & 18.47$\pm 0.06$ & \nodata         & 17.38$\pm 0.08$ & \nodata         & clear \nl
 Feb 11.22  &  52681.22  & \nodata         & \nodata         & \nodata         & \nodata         & \nodata         & \nodata         & $<$ 19.7        & clear \nl
 Feb 23.22  &  52693.22  & 20.99$\pm 0.42$ & \nodata         & \nodata         & \nodata         & \nodata         & \nodata         & \nodata         & clear \nl
 Feb 24.23  &  52694.23  & \nodata         & 20.49$\pm 0.11$ & \nodata         & \nodata         & \nodata         & \nodata         & \nodata         & clear \nl
 Feb 25.23  &  52695.23  & \nodata         & \nodata         & 19.60$\pm 0.06$ & \nodata         & \nodata         & \nodata         & \nodata         & clear \nl
 Feb 26.23  &  52696.23  & \nodata         & \nodata         & \nodata         & 19.50$\pm 0.12$ & \nodata         & \nodata         & \nodata         & clear \nl
 Mar 11.21  &  52709.21  & \nodata         & \nodata         & \nodata         & \nodata         & \nodata         & 17.89$\pm 0.13$ & \nodata         & clear \nl
 Mar 12.22  &  52710.22  & \nodata         & \nodata         & \nodata         & \nodata         & 19.16$\pm 0.22$ & \nodata         & \nodata         & clear \nl
 Jun 27.62  &  52817.62  & \nodata         & \nodata         & 21.26$\pm 0.22$ & \nodata         & \nodata         & \nodata         & \nodata         & clear \nl
 Jul  3.60  &  52823.60  & 22.80$\pm 1.36$ & \nodata         & \nodata         & \nodata         & \nodata         & \nodata         & \nodata         & clear \nl
 Jul  5.60  &  52825.60  & \nodata         & 21.75$\pm 0.31$ & \nodata         & \nodata         & \nodata         & 18.80$\pm 0.21$ & \nodata         & clear \nl
 Aug 15.49  &  52866.49  & \nodata         & \nodata         & \nodata         & $<$ 21.5        & \nodata         & \nodata         & \nodata         & thin cloud \nl
 Aug 28.50  &  52879.50  & \nodata         & \nodata         & \nodata         & \nodata         & $<$ 21.2        & 19.83$\pm 0.34$ & \nodata         & clear \nl
 Aug 29.50  &  52880.50  & \nodata         & \nodata         & \nodata         & \nodata         & \nodata         & \nodata         & $<$ 19.9        & clear \nl
 \enddata
\end{deluxetable}

\begin{deluxetable}{lcccccccc}
 \rotate
 \tabletypesize{\scriptsize}
 \tablecaption{Decline rates of light curves of SN~2002ap in the nebular phase}
 \tablewidth{0pt}
 \tablenum{2}
 \tablehead{
 \colhead{}          &
 \colhead{$B$ band}  & \colhead{$V$ band}  & \colhead{$R$ band}  & \colhead{$I$ band}  &
 \colhead{$J$ band}  & \colhead{$H$ band}  & \colhead{$K$ band}  & \colhead{Bolometric}
}
 \startdata
 Epoch range (MJD) &
 52438.6 - 52533.5 & 52437.6 - 52533.5 & 52439.6 - 52817.6 & 52441.6 - 52696.2 &
 52439.6 - 52508.5 & 52438.6 - 52524.6 & 52437.6 - 52577.5 & 52438.5 - 52533.5 \nl

 Decline rate (mag/day) &
 0.017$\pm 0.0004$ & 0.021$\pm 0.0002$ & 0.016$\pm 0.0001$ & 0.017$\pm 0.0001$ &
 0.018$\pm 0.001$ & 0.018$\pm 0.001$   & 0.022$\pm 0.001$  & 0.018$\pm 0.0003$ \nl

\nl

 Epoch range (MJD) &
 52574.5 - 52823.6 & 52574.5 - 52825.6 & \nodata & \nodata           &
 52522.6 - 52710.2 & 52533.5 - 52879.5 & \nodata & 52574.5 - 52879.5 \nl

 Decline rate (mag/day) &
 0.015$\pm 0.002$ & 0.014$\pm 0.001$  & \nodata  & \nodata          &
 0.012$\pm 0.001$ & 0.008$\pm 0.0005$ & \nodata  & 0.014$\pm 0.001$ \nl

\nl

 Date of change (MJD) &
 52545$\pm 5$ & 52535$\pm 1$ & \nodata & \nodata      &
 52514$\pm 4$ & 52532$\pm 4$ & \nodata & 52558$\pm 3$

 \enddata
\end{deluxetable}

\begin{deluxetable}{lccc}
 \tablecaption{Bolometric magnitudes of SN~2002ap}
 \tablewidth{0pt}
 \tablenum{3}
 \tablehead{
 \colhead{MJD}\tablenotemark{a} & \colhead{Day}\tablenotemark{b} & \colhead{$M_{\rm bol}$} & \colhead{$L_{\rm NIR}/L_{\rm bol}$}
}
 \startdata
 52438.5  & 136.5 & -13.53$\pm$0.03 & 23\% \nl
 52458.5  & 156.5 & -13.17$\pm$0.03 & 22\% \nl
 52478.5  & 176.5 & -12.81$\pm$0.03 & 22\% \nl
 52493.5  & 191.5 & -12.52$\pm$0.03 & 22\% \nl
 52504.0  & 202.0 & -12.32$\pm$0.03 & 21\% \nl
 52508.5  & 206.5 & -12.28$\pm$0.03 & 23\% \nl
 52521.0  & 219.0 & -12.04$\pm$0.05 & 21\% \nl
 52525.0  & 223.0 & -11.96$\pm$0.03 & 21\% \nl
 52533.5  & 231.5 & -11.83$\pm$0.03 & 22\% \nl
 52574.5  & 272.5 & -11.15$\pm$0.05 & 23\% \nl
 52577.5  & 275.5 & -11.14$\pm$0.05 & 24\% \nl
 52601.0  & 299.0 & -10.74$\pm$0.07 & 25\% \nl
 52609.5  & 307.5 & -10.62$\pm$0.07 & 26\% \nl
 52617.5  & 315.5 & -10.50$\pm$0.07 & 26\% \nl
 52626.5  & 324.5 & -10.37$\pm$0.09 & 29\% \nl
 52638.5  & 336.5 & -10.23$\pm$0.09 & 29\% \nl
 52695.0  & 393.0 & -9.47$_{-0.18}^{+0.17}$ & 39\% \nl
 52709.0  & 407.0 & -9.3$\pm$ 0.2 & \nodata \nl
 52822.0  & 520.0 & -8.06$_{-0.43}^{+0.30}$ & 54\% \nl
 52879.5  & 577.5 & -6.8$\pm$ 0.5 & \nodata
 \enddata

 \tablenotetext{a}{With an accuracy of 0.5 days due to data interpolation/extrapolation.}

 \tablenotetext{b}{With respect to the explosion date (MJD 52302.0)}
\end{deluxetable}

\begin{deluxetable}{cccccc}
 \rotate
 \tabletypesize{\scriptsize}
 \tablecaption{Type Ib/c supernovae with late-time NIR photometry}
 \tablewidth{0pt}
 \tablenum{4}
 \tablehead{
 \colhead{} & \colhead{SN 1982R} & \colhead{SN 1983I} & \colhead{SN 1983N} & \colhead{SN 1984L} & \colhead{SN 1998bw}
}
 \startdata
 Spectroscopic type                    & Ib/c & Ic   & Ib    & Ib   & Ic, hypernova \nl
 Distance modulus \tablenotemark{a}    & 31.2 & 31.0 & 28.25 & 31.5 & 32.76 \nl
 Extinction $A(V)$ \tablenotemark{b}  & 0.5 & 0.04 & 0.2  & 0.09 & 0.1$\pm$0.1 \nl
 Explosion date (UT) \tablenotemark{c} & 1982 September 28 & 1983 April 25 & 1983 July 1 & 1984 August 5 & 1998 April 25.9 \nl

\nl

 JHK photometric &               &               & \citet{eli85}, &               & \citet{pat01}, \nl
 data sources    & \citet{mat01} & \citet{eli85} & \citet{mat01}  & \citet{eli85} & \citet{sol02}

 \enddata
\tablenotetext{a}{$h_0=0.72$, $\Omega_\Lambda=0.7$,
$\Omega_M$=0.3; 1982R \& 1984L, radial velocity corrected for LG
infall onto Virgo (LEDA; \citealt{pat97}); 1983I, see \S 3.4;
1983N, Cepheids \citep{thi03}; 1998bw, z=0.0085 \citep{gal98}.}

\tablenotetext{b}{1982R, \citet{clo96}; 1983I, 1983N, \& 1984L,
Galactic \citet{sch98}; 1998bw, \citet{pat01}.}

\tablenotetext{c}{1982R, 26 days before discovery \citep{por87};
1983I, see \S 3.4; 1983N, 2 days before discovery and 16 days
before B maximum \citep{por87}; 1984L, 15 days before light
maximum and by comparison with SN 1985F \citep{tsv87,sch89};
1998bw, GRB 980425 \citep{gal98}.}

\end{deluxetable}

}

%
\end{document}